\documentclass[aps,preprint,tighten,showpacs,graphicx]{revtex4}

\usepackage[dvips]{graphicx}
\newcommand{\la}{\left\langle}
\newcommand{\ra}{\right\rangle}

\newcommand{\PRL}{{\it Phys.~Rev.~Lett.~}}
\newcommand{\PR}{{\it Phys.~Rev.~}}
\newcommand{\JCP}{{\it J.~Chem.~Phys.~}}
\newcommand{\JPC}{{\it J.~Phys.~Chem.~}}
\newcommand{\JPCM}{{\it J.~Phys.: Condens.~Matter~}}
\newcommand{\MP}{{\it Mol.~Phys.~}}

\begin{document}

\title{Effective Electrostatic Interactions in Suspensions of 
Polyelectrolyte Brush-Coated Colloids}

\author{H. Wang} \email{hao.wang@ndsu.nodak.edu}
\author{A. R. Denton} \email{alan.denton@ndsu.nodak.edu}
\affiliation{Department of Physics, North Dakota State University,
Fargo, ND, 58105-5566}

\date{\today}

\begin{abstract}
Effective electrostatic interactions between colloidal particles, coated 
with polyelectrolyte brushes and suspended in an electrolyte solvent, 
are described via linear response theory.  The inner cores of the macroions
are modeled as hard spheres, the outer brushes as spherical shells of 
continuously distributed charge, the microions (counterions and salt ions) 
as point charges, and the solvent as a dielectric continuum.  
The multi-component mixture of macroions and microions is formally mapped 
onto an equivalent one-component suspension by integrating out from 
the partition function the microion degrees of freedom.
Applying second-order perturbation theory and a random phase approximation,
analytical expressions are derived for the effective pair interaction 
and a one-body volume energy, which is a natural by-product of the 
one-component reduction.  The combination of an inner core and an 
outer shell, respectively impenetrable and penetrable to microions, 
allows the interactions between macroions to be tuned by varying the 
core diameter and brush thickness.  In the limiting cases of vanishing 
core diameter and vanishing shell thickness, the interactions reduce to 
those derived previously for star polyelectrolytes and charged colloids, 
respectively.
\end{abstract}


\maketitle

\section{Introduction}

Polyelectrolytes~\cite{PE1,PE2} are ionizable polymers that dissolve in 
a polar solvent, such as water, through dissociation of counterions.  
Solutions of polyelectrolytes are complex mixtures of macroions and microions 
(counterions and salt ions) in which direct electrostatic interactions 
between macroions are screened by surrounding microions.
Polyelectrolyte chains, grafted or adsorbed by one end to a surface 
at high concentration, form a dense brush that can significantly modify 
interactions between surfaces in solution.
When attached to colloidal particles, {\it e.g.}, latex particles in paints 
or casein micelles in milk~\cite{Tuinier02}, polyelectrolyte brushes 
can stabilize colloidal suspensions by inhibiting 
flocculation~\cite{Evans,Hunter}.
Biological polyelectrolytes (biopolymers), such as proteins in 
cell membranes, can modify intercellular and cell-surface interactions.

Conformations and density profiles of polyelectrolyte (PE) brushes have been 
studied by a variety of experimental, theoretical, and simulation methods, 
including dynamic light scattering~\cite{Guo-Ballauff01}, 
small-angle neutron scattering~\cite{Mir95,Guenoun98,Groenewegen00}, 
transmission electron microscopy~\cite{Groenewegen00},
neutron reflectometry~\cite{Tran99}, 
surface adsorption~\cite{Hariharan-Russel98}, 
atomic force microscopy~\cite{Mei-Ballauff03},
self-consistent field theory~\cite{Miklavic88,Misra89,Misra96,
Zhulina-Borisov97,Gurovitch-Sens99,Borisov01,Klein-Wolterink-Borisov03}, 
scaling theory~\cite{Pincus91,Schiessel-Pincus98,Borisov01,
Klein-Wolterink-Borisov03}, 
Poisson-Boltzmann theory~\cite{Miklavic90}, 
Monte Carlo simulation~\cite{Miklavic90}, and
molecular dynamics simulation~\cite{Seidel00,Likos02}.
Comparatively few studies have focused on electrostatic interactions between 
PE brush-coated surfaces.  Interactions between neutral surfaces 
-- both planar and curved (spherical) -- with grafted PE brushes 
have been modeled using scaling theory~\cite{Pincus91}, while
interactions between charged surfaces coated with oppositely-charged 
PEs have been investigated for planar~\cite{Miklavic90} and spherical 
(colloidal) surfaces~\cite{Podgornik95} via Monte Carlo simulation 
and a variety of theoretical methods.

While microscopic models that include chain and microion degrees of freedom 
provide the most realistic description of PE brushes, simulation of such 
explicit models for more than one or two brushes can be computationally 
demanding. 
The purpose of the present paper is to develop an alternative, 
coarse-grained theoretical approach, based on the concept of effective 
interactions, which may prove useful for predicting thermodynamic and 
other bulk properties of suspensions of PE brush-coated colloids.  
Modeling each brush as a spherical shell of continuously distributed 
charge, we adapt linear response theory, previously developed for charged 
colloids~\cite{Silbert91,Denton99,Denton00} and PEs~\cite{Denton03}, 
to derive effective electrostatic interactions.  
The theory is based on mapping the multi-component mixture 
onto an equivalent one-component system of ``pseudo-macroions" by 
integrating out from the partition function the degrees of freedom 
of the microions. 
Within the theory, microions play three physically important roles: 
reducing (renormalizing) the bare charge on a macroion; 
screening direct Coulomb interactions between macroions; 
and generating a one-body volume energy.  
The volume energy -- a natural by-product of the one-component reduction 
-- contributes to the total free energy and can significantly influence 
thermodynamic behavior of deionized suspensions.

Outlining the remainder of the paper, Sec.~\ref{Model} defines the
model suspension of PE brush-coated colloids; Sec.~\ref{Theory} reviews
the linear response theory; Secs.~\ref{Analytical Results} 
and \ref{Numerical Results} present analytical and numerical results 
for counterion density profiles, effective pair interactions, 
and volume energies in bulk suspensions;
and finally, Sec.~\ref{Conclusions} summarizes and concludes.

\section{Model}\label{Model}

\begin{figure}
\includegraphics[width=0.6\columnwidth]{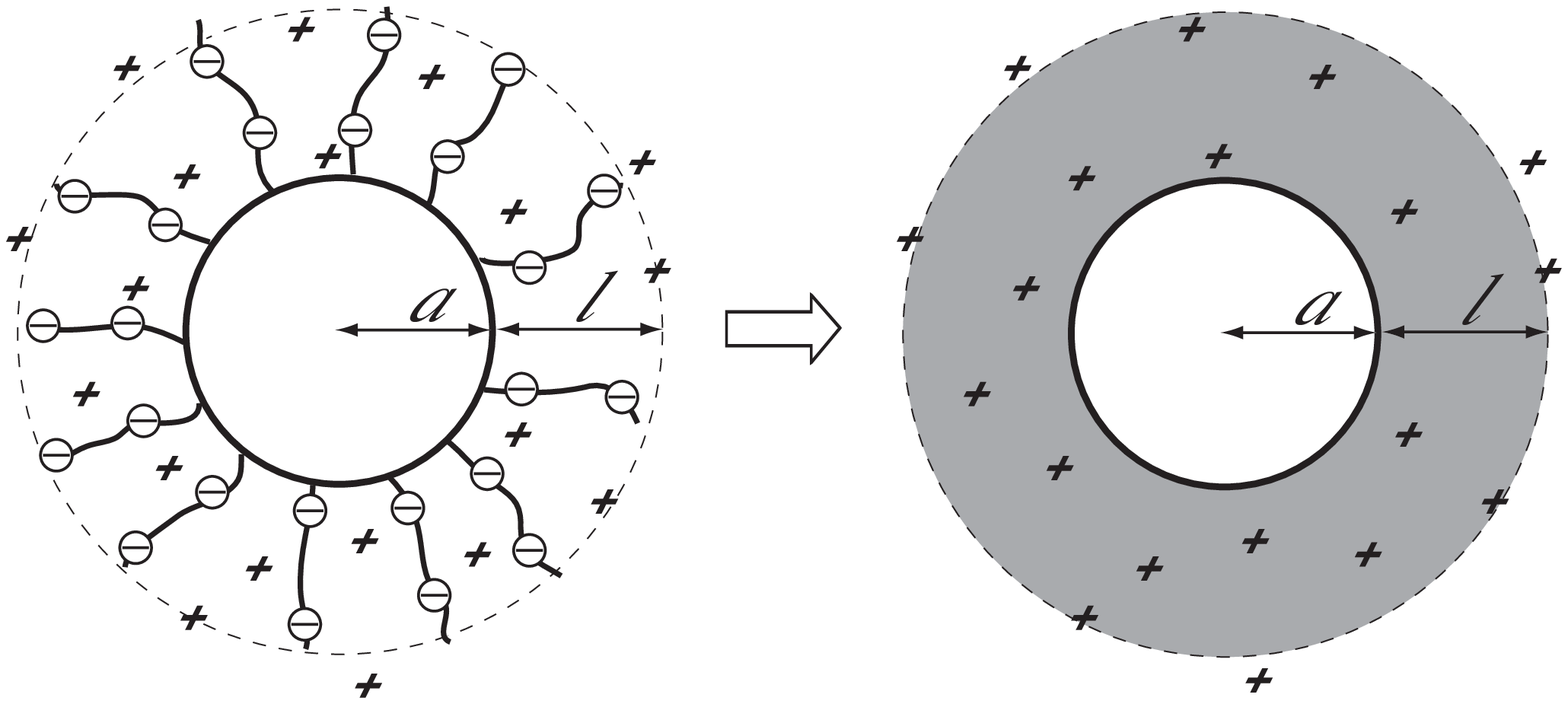}
\caption{\label{PEbrush} (a) Polyelectrolyte (PE) brush-coated
colloidal sphere and, (b) model considered here, in which the 
PE monomer charge distribution is assumed continuous and varying 
as $1/r^2$, $a<r<a+l$.}
\end{figure}

The system of interest is modeled as a suspension of $N_m$ spherical, 
core-shell macroions of charge $-Ze$ (valence $Z$), core radius $a$, 
and PE brush shell thickness $l$ (outer radius $R=a+l$), and
$N_c$ point counterions of charge $ze$ in an electrolyte solvent 
in volume $V$ at temperature $T$ (see Fig.~\ref{PEbrush}).  
The core is assumed to be neutral, the macroion charge coming entirely
from the PE shell. 
Assuming a symmetric electrolyte and equal salt and counterion valences, 
the electrolyte contains $N_s$ point salt ions of charge $ze$ and 
$N_s$ of charge $-ze$.  The microions thus 
number $N_+=N_c+N_s$ positive and $N_-=N_s$ negative, for a total of
$N_{\mu}=N_c+2N_s$.  Global charge neutrality in a bulk suspension
constrains macroion and counterion numbers via $ZN_m=zN_c$.  
Number densities of macroions, counterions, and salt ions are denoted 
by $n_m$, $n_c$, and $n_s$, respectively.
Within the primitive model of ionic liquids~\cite{HM}, the solvent 
is treated as a dielectric continuum of dielectric constant $\epsilon$, 
which acts only to reduce the strength of Coulomb interactions between ions.

In PE solutions, the counterions can be classified into four regions:
(1) those within narrow tubes enclosing the PE chains, of radius 
comparable to the Bjerrum length, $\lambda_B=e^2/(\epsilon k_{\rm B}T)$;
(2) those outside of the tubes but still closely associated with the chains; 
(3) those not closely associated with the chains, but still inside of 
the PE shells; and (4) those entirely outside of the macroions.
Counterions in regions (1)-(3) can be regarded as trapped by the macroions,
while those in region (4) are free to move throughout the suspension.
Within region (1), the counterions may be either condensed and immobilized 
on a chain or more loosely bound and free to move along a chain.
These chain-localized (condensed or mobile) counterions tend to distribute 
uniformly along, and partially neutralize, the chains.
In our model, counterions in regions (1) and (2) act to renormalize the 
bare macroion valence.  The parameter $Z$ thus should be physically
interpreted as an {\it effective} macroion valence, generally
much lower than the bare valence (number of ionizable monomers).
From the Manning counterion condensation criterion~\cite{PE1}, according 
to which the linear charge density of a PE chain saturates at 
$\sim e/\lambda_B$, we can expect the bare charge in an aqueous solution 
to be renormalized down by at least an order of magnitude.

The local number density profiles of charged monomers in the PE brushes, 
$\rho_{\rm mon}(r)$, and of counterions, $\rho_c(r)$, are modeled here
as continuous, spherically symmetric distributions.  Charge discreteness 
can be reasonably neglected if we ignore structure on length scales 
shorter than the minimum separation between charges.  Spherical symmetry 
of charge distributions can be assumed if intra-macroion chain-chain 
interactions, which favor isotropic distribution of chains, dominate 
over inter-macroion interactions, which favor anisotropy.

The density profile of charged monomers depends on the conformations 
of chains in the PE shells.  Electrostatic repulsion between charged 
monomers tends to radially stretch and stiffen PE chains. 
Indeed, neutron scattering experiments~\cite{Guenoun98} on diblock 
(neutral-charged) copolymer micelles, as well as simulations~\cite{Likos02},
provide strong evidence that the arms of spherical PE brushes 
can exhibit rodlike behavior.
Here we assume the ideal case of fully stretched chains of equal length
-- a porcupine conformation~\cite{Pincus91} -- and model the 
charged monomer number density profile by
\begin{equation}
\rho_{\rm mon}(r)~=~~\left\{
\begin{array} {l@{\quad\quad}l}
~~~0, & r>R \\
{\displaystyle \frac{Z}{4\pi lr^2}}, & a < r \leq R\\
~~~0, & r\leq a, \end{array} \right .\label{monomerdensity}
\end{equation}
where $r$ is the radial distance from the macroion's center.
The model thus neglects configurational entropy of the PE chains, 
although it does include the entropy of the microions.

\section{Theory}\label{Theory}

For the model suspension defined above, our goal is to predict 
distributions of microions inside and outside of the PE brushes
and effective interactions between macroions.
Adapting the general response theory approach previously applied 
to charged colloids~\cite{Silbert91,Denton99,Denton00} 
and PE solutions~\cite{Denton03}, we reduce the multi-component mixture 
to an equivalent one-component system governed by effective interactions, 
and approximate the effective one-component Hamiltonian via 
perturbation theory.  To simplify notation, we initially ignore salt ions. 
The Hamiltonian then decomposes, quite generally, into three terms:
\begin{equation}
H=H_m(\{{\bf R}\})+H_c(\{{\bf r}\})+H_{mc}(\{{\bf R}\},\{{\bf r}\}), 
\label{H}
\end{equation}
where $\{{\bf R}\}$ and $\{{\bf r}\}$ denote collective coordinates 
of macroions and counterions, respectively.
The first term in Eq.~(\ref{H}),
\begin{equation}
H_m=H_{\rm hc}+\frac{1}{2}\sum_{i\neq j=1}^{N_m}v_{mm}(|{\bf R}_i-{\bf
R}_j|), \label{Hp}
\end{equation}
is the macroion Hamiltonian, which includes a hard-core contribution
$H_{\rm hc}$ (kinetic energy and hard-core interactions), and an
electrostatic contribution due to the bare Coulomb pair interaction potential 
\begin{equation}
v_{mm}(r)=\frac{Z^2e^2}{\epsilon r}
\label{vmm}
\end{equation}
at center-center separation $r$.
The second term in Eq.~(\ref{H}),
\begin{equation}
H_c=K_c+\frac{1}{2}\sum_{i\neq j=1}^{N_c}v_{cc}(|{\bf r}_i-{\bf r}_j|), 
\label{Hc}
\end{equation}
is the Hamiltonian of the counterions with kinetic energy $K_c$ interacting 
via the Coulomb pair potential $v_{cc}(r)=z^2e^2/\epsilon r$. 
The third term in Eq.~(\ref{H}),
\begin{equation}
H_{mc}=\sum_{i=1}^{N_m}\sum_{j=1}^{N_c}v_{mc}(|{\bf R}_i-{\bf r}_j|), 
\label{Hmc1}
\end{equation}
is the macroion-counterion interaction, which also may be expressed in the form
\begin{equation}
H_{mc}~=~\int{\rm d}{\bf R}\,\rho_m({\bf R})\int{\rm d}{\bf r}
\,\rho_c({\bf r})v_{mc}(|{\bf R}-{\bf r}|), \label{Hmc2}
\end{equation}
where $\rho_m({\bf R})=\sum_{i=1}^{N_m} \delta({\bf R}-{\bf R}_i)$
and $\rho_c({\bf r})=\sum_{j=1}^{N_c} \delta({\bf r}-{\bf r}_j)$
are the macroion and counterion number density operators, respectively,
and $v_{mc}(|{\bf R}-{\bf r}|)$ is the macroion-counterion interaction
potential (to be specified in Sec.~\ref{Analytical Results}).

The mixture of macroions and counterions can be formally reduced to an
equivalent one-component system by integrating out the counterion coordinates. 
Denoting traces over counterion and macroion coordinates by 
$\la~\ra_c$ and $\la~\ra_m$, respectively, the canonical partition function 
can be expressed as
\begin{equation}
{\cal Z}~=~\la\la\exp(-\beta H)\ra_c\ra_m
~=~\la\exp(-\beta H_{\rm eff})\ra_m, \label{part}
\end{equation}
where $\beta=1/k_BT$, $H_{\rm eff}=H_m+F_c$ is the effective 
one-component Hamiltonian, and
\begin{equation}
F_c~=~-k_BT\ln\la\exp\Bigl[-\beta(H_c+H_{mc})\Bigr]\ra_c
\label{Fc1}
\end{equation}
is the free energy of a nonuniform gas of counterions in the
presence of the macroions.

Now regarding the macroions as an ``external" potential for the counterions,
we invoke perturbation theory~\cite{Silbert91,Denton99,Denton00,HM} and write
\begin{equation}
F_c~=~F_0~+~\int_0^1{\rm d}\lambda\,\la H_{mc}\ra_{\lambda}, \label{Fc2}
\end{equation}
where $F_0=-k_BT\ln\la\exp(-\beta H_c)\ra_c$ is the reference free energy of
the unperturbed counterions, the $\lambda$-integral charges the macroions
({\it i.e.}, the PE brushes) from neutral to fully charged,
$H_{mc}$ represents the perturbing potential of the macroions acting 
on the counterions, and $\la H_{mc}\ra_{\lambda}$ is the mean value 
of this potential in a suspension of macroions charged to a fraction 
$\lambda$ of their full charge.

Two formal manipulations prove convenient.  First, we convert the 
free energy of the unperturbed counterions to that of a classical 
one-component plasma (OCP) by adding and subtracting, on the right side 
of Eq.~(\ref{Fc2}), the energy of a uniform compensating negative 
background~\cite{note1}, $E_b=-N_cn_c\hat v_{cc}(0)/2$.  
Here $n_c=N_c/[V(1-\eta_{\rm hc})]$ is the average density of 
counterions in the free volume -- {\it i.e.}, the total volume
reduced by the volume fraction $\eta_{\rm hc}=(4\pi/3)n_ma^3$ of 
the macroion hard cores -- and $\hat v_{cc}(0)$ 
is the $k \to 0$ limit of the Fourier transform of $v_{cc}(r)$.  
Equation~(\ref{Fc2}) then becomes
\begin{equation}
F_c~=~F_{\rm OCP}~+~\int_0^1{\rm d}\lambda\,\la H_{mc}\ra_{\lambda}
-E_b, \label{Fc3}
\end{equation}
where $F_{\rm OCP}=F_0+E_b$ is the free energy of a homogeneous OCP
excluded from the colloidal hard cores.
Second, we express $H_{mc}$ in terms of Fourier components:
\begin{equation}
\la H_{mc}\ra_{\lambda}~=~\frac{1}{V}\sum_{{\bf k}\neq 0} \hat
v_{mc}(k) \hat\rho_m({\bf k}) \la\hat\rho_c(-{\bf k})\ra_{\lambda}
+ \frac{1}{V}\lim_{k\to 0}\left[\hat v_{mc}(k) \hat\rho_m({\bf k})
\la\hat\rho_c(-{\bf k})\ra_{\lambda}\right], \label{Hmck}
\end{equation}
where $\hat v_{mc}(k)$ is the Fourier transform of the
macroion-counterion interaction and where
$\hat\rho_m({\bf k})=\sum_{j=1}^{N_m}\exp(-i{\bf k}\cdot{\bf R}_j)$
and $\hat\rho_c({\bf k})=\sum_{j=1}^{N_c}\exp(-i{\bf k}\cdot{\bf
r}_j)$ are Fourier components of the macroion and counterion number 
density operators.  The $k=0$ term is singled out because the number 
of counterions, $N_c=\hat\rho_c(0)$, does not respond to the macroion 
charge, but rather is fixed by the constraint of global charge neutrality.

Further progress requires approximations for the counterion free energy.
Applying second-order perturbation (linear response) theory, the counterions
are assumed to respond linearly to the macroion external potential:
\begin{equation}
\rho_c({\bf r})~=~\int{\rm d}{\bf r}'\,\chi({\bf r}-{\bf r}')
\int{\rm d}{\bf r}''\,\rho_m({\bf r}'')v_{mc}({\bf r}'-{\bf r}'')
\label{rhocr}
\end{equation}
or
\begin{equation}
\hat\rho_c({\bf k})~=~\chi(k)\hat v_{mc}(k)\hat\rho_m({\bf k}),
\qquad k\neq 0, \label{rhock}
\end{equation}
where $\chi(k)$ is the linear response function of the OCP.

Combining Eqs.~(\ref{Fc3})-(\ref{rhock}), the effective Hamiltonian 
can be expressed in the form of the Hamiltonian of a one-component 
pairwise-interacting system:
\begin{equation}
H_{\rm eff}~=~H_{\rm hc}~+~\frac{1}{2}\sum_{i\neq j=1}^{N_m}
v_{\rm eff}(|{\bf R}_i-{\bf R}_j|)~+~E_0, \label{Heff}
\end{equation}
where $v_{\rm eff}(r)=v_{mm}(r)+v_{\rm ind}(r)$ is an effective
electrosatic macroion-macroion pair interaction that augments the 
bare macroion interaction $v_{mm}(r)$ by a counterion-induced interaction
\begin{equation}
\hat v_{\rm ind}(k)~=~\chi(k)\left[\hat v_{mc}(k)\right]^2.
\label{vindk}
\end{equation}
The final term in Eq.~(\ref{Heff}) is the volume energy, 
\begin{equation}
E_0~=~F_{\rm OCP}~+~\frac{N_m}{2}\lim_{r\to 0} v_{\rm
ind}(r)~+~N_m\lim_{k\to 0}\left[-\frac{1}{2}n_m\hat v_{\rm
ind}(k)+n_c\hat v_{mc}(k)+\frac{Z}{2z}n_c\hat v_{cc}(k)\right],
\label{E0}
\end{equation}
which emerges naturally from the one-component reduction.
Although independent of the macroion coordinates, the volume energy 
depends on the average macroion density and thus has the potential
to significantly influence thermodynamics.
Evidently, the effective interactions depend on the macroion structure
through the specific form of the macroion-counterion interaction $v_{mc}$
in Eqs.~(\ref{vindk}) and (\ref{E0}).

The OCP linear response function, proportional to the corresponding
static structure factor $S(k)$, may be obtained from
liquid-state theory~\cite{HM}.  In practice, the OCP is weakly
correlated, with coupling parameter $\Gamma=\lambda_B/a_c \ll 1$,
where $a_c=(3/4\pi n_c)^{1/3}$ is the counterion sphere radius. 
For example, for hard-sphere macroions of radius $a=50$ nm, 
valence $Z=500$, and volume fraction $\eta_{\rm hc}=0.01$, in water at 
room temperature ($\lambda_B=0.714$ nm), we find $\Gamma\simeq 0.02$. 
As in previous work on charged colloids~\cite{Denton99,Denton00}
and polyelectrolytes~\cite{Denton03}, we adopt the random phase 
approximation (RPA), which is valid for weakly-coupled plasmas. 
The RPA equates the OCP two-particle direct correlation function 
to its exact asymptotic limit: $c^{(2)}(r)=-\beta v_{cc}(r)$. 
Using the Ornstein-Zernike relation, $S(k)=1/[1-n_c\hat c^{(2)}(k)]$, 
the linear response function then takes the form
\begin{equation}
\chi(k)~=~-\beta n_c S(k)~=~-\frac{\beta n_c}{(1+\kappa^2/k^2)},
\label{chi}
\end{equation}
where $\kappa=\sqrt{4\pi n_cz^2\lambda_B}$ is the Debye screening
constant (inverse screening length).  Note that the screening constant,
which involves the density of counterions in the free volume,
naturally incorporates the excluded volume of the macroion cores.
With $\chi(k)$ specified, the counterion density can be calculated from 
the macroion-counterion interaction and Eq.~(\ref{rhock}) for a given
macroion distribution (see Sec.~\ref{Analytical Results}).  Finally, 
salt ions can be easily incorporated by introducing additional 
response functions~\cite{Denton00}.  The pair interaction and 
volume energy are then modified only through a redefinition of 
the Debye screening constant:
$\kappa=\sqrt{4\pi (n_c+2n_s)z^2\lambda_B}$, where 
$n_s=N_s/[V(1-\eta_{\rm hc})]$ is the average number density 
of salt ion pairs in the free volume.

Generalization of response theory to incorporate leading-order 
nonlinear microion response entails three-body effective interactions, 
as well as corrections to the effective pair potential and 
volume energy~\cite{Denton04}.  Nonlinear effects are generally 
significant, however, only in concentrated, deionized suspensions 
of highly charged macroions~\cite{Denton04} and are here ignored. 
It has been shown that response theory, combined with the RPA, 
is formally equivalent to Poisson-Boltzmann theory~\cite{Denton04}.
Both approaches rely on mean-field approximations that neglect 
microion fluctuations and predict microion distributions of the 
same general form, aside from a distinction in the screening constant, 
which response theory corrects for excluded volume of the macroion cores. 
Advantages of response theory over Poisson-Boltzmann theory are its
predictions of (1) the entire effective Hamiltonian, including the 
one-body volume energy, which is essential for a complete description 
of phase behavior~\cite{Denton99,Denton00,Silbert91,vRH,Graf,vRDH,Warren},
and (2) a more accurate expression for the Debye screening constant
that incorporates the macroion excluded-volume correction.

\section{Analytical Results}\label{Analytical Results}

For our porcupine model of a spherical PE brush with $1/r^2$ monomer 
density profile, Gauss's law gives the electric field around a macroion as
\begin{equation}
E(r)~=~\left\{ \begin{array}
{l@{\quad\quad}l}
-\frac{\displaystyle Ze}{\displaystyle \epsilon}\frac{\displaystyle 1}{\displaystyle r^2}, & r>R \\
-\frac{\displaystyle Ze}{\displaystyle
\epsilon}\frac{\displaystyle r-a}{\displaystyle lr^2}, & a < r \leq R\\
~~~~ 0 , & r\leq a.
\end{array} \right. \label{Estar}
\end{equation}
Integration over $r$ yields the electrostatic potential energy
between a brush and a counterion:
\begin{equation}
v_{mc}(r)~=~\left\{ \begin{array} {l@{\quad\quad}l}
~~~~~-\frac{\displaystyle Zze^2}{\displaystyle \epsilon r}, & r>R \\
-\frac{\displaystyle Zze^2}{\displaystyle \epsilon
l}\left[1-\frac{\displaystyle a}{\displaystyle
r}-\ln\left(\frac{\displaystyle r}{\displaystyle
R}\right)\right], & a < r \leq R\\
-\frac{\displaystyle Zze^2}{\displaystyle \epsilon
l}\left[\alpha-\ln{\left(\frac{\displaystyle a}{\displaystyle
R}\right)}\right], & r \leq a,
\end{array} \right. \label{vmcrbrush}
\end{equation}
where $\alpha$ is an arbitrary constant, which arises because 
$v_{mc}(r)$ is not uniquely defined inside the hard core.
Following van Roij and Hansen~\cite{vRDH}, we choose $\alpha$ below
by requiring that the counterion density vanish inside the hard core. 
Fourier transforming Eq.~(\ref{vmcrbrush}) yields
\begin{equation}
\hat v_{mc}(k)~=~ -\frac{4\pi Zze^2}{\epsilon k^3 l}{\rm G}(ka,kR;\alpha), 
\label{vmckbrush}
\end{equation}
where the function ${\rm G}(ka,kR;\alpha)$ is defined as
\begin{equation}
{\rm G}(x_{1},x_{2};\alpha)~=~{\rm sinc}(x_{2})-{\rm
sinc}(x_{1})-\alpha[x_{1}\cos(x_{1})-\sin(x_{1})],
\label{functionG}
\end{equation}
with ${\rm sinc}(x)\equiv\int_0^x{\rm d}u\,\sin(u)/u$. 
We can now calculate, in the dilute limit, the counterion number density 
profile around a single macroion, taking $\hat\rho_m({\bf k})=1$.  
From Eqs.~(\ref{rhock}), (\ref{chi}), and
(\ref{vmckbrush}), the Fourier component of the density profile is 
\begin{equation}
\hat\rho_c(k)~=~\frac{Z}{z}\frac{\kappa^2}{kl(k^2+\kappa^2)}
{\rm G}(ka,kR;\alpha), \label{rhockbrush}
\end{equation}
which in real space takes the form 
\begin{equation}
\rho_c(r)~=~\frac{Z}{z}\frac{\kappa}{4\pi lr}~\left\{
\begin{array} {l@{\quad\quad}l}
{\rm S}(\kappa a,\kappa R;\alpha)~e^{-\kappa r}, & r>R \\
{\rm S}(\kappa a,\kappa r;\alpha)~e^{-\kappa r}
+{\rm Ec}(-\kappa r,-\kappa R){\rm sinh}(\kappa r), & a < r \leq R\\
\left[{\rm Ec}(-\kappa a,-\kappa R)+\alpha(1+\kappa a)e^{-\kappa
a}\right]{\rm sinh}(\kappa r), & r \leq a.
\end{array} \right. \label{rhocrbrushfull}
\end{equation}
For simplicity, we have introduced two functions,
${\rm S}(x_{1},x_{2};\alpha)$ and ${\rm Ec}(x_{1},x_{2})$, which are
defined, respectively, as
\begin{equation}
{\rm S}(x_{1},x_{2};\alpha)={\rm shi}(x_{2})-{\rm
shi}(x_{1})-\alpha[x_{1}\cosh(x_{1})-\sinh(x_{1})]
\label{functionS}
\end{equation}
and
\begin{equation}
{\rm Ec}(x_{1},x_{2})={\rm Ei}(x_{2})-{\rm Ei}(x_{1}),
\label{findAlan}
\end{equation}
where ${\rm shi}(x)\equiv\int_0^x{\rm d}u\,\sinh(u)/u$ 
denotes the hyperbolic sine integral function and
${\rm Ei}(x)\equiv\int_{-\infty}^{x}{\rm d}u\,e^u/u$ 
is the exponential integral function.
Now setting the counterion density to zero within the hard core
[{\it i.e.}, $\rho_c(r)=0, r\leq a$, in Eq.~(\ref{rhocrbrushfull})]
fixes the constant $\alpha$:
\begin{equation}
\alpha~=~-\frac{e^{\kappa a}}{1+\kappa a}{\rm Ec}(-\kappa a,
-\kappa R). \label{constant}
\end{equation}

Integrating Eq.~(\ref{rhocrbrushfull}) over the spherical shell volume 
of a PE brush yields the fraction of counterions inside a brush:
\begin{equation}
f_{\rm in}~=~1-\frac{1+\kappa R}{\kappa l}~e^{\displaystyle
-\kappa R}~{\rm S}(\kappa a,\kappa R;\alpha).
\label{ncinside-brush}
\end{equation}
From this expression, it is clear that the counterion distribution, 
within the model, is determined entirely by two independent dimensionless 
parameters, $\kappa a$ and $\kappa l$, {\it i.e.}, the ratios of the macroion 
core radius and brush thickness, respectively, to the Debye screening length.

From Eqs.~(\ref{vindk}) and (\ref{vmckbrush}), the induced
electrostatic pair interaction is given by
\begin{equation}
\hat v_{\rm ind}(k) ~=~-\frac{4\pi
Z^2e^2}{\epsilon}~\frac{\kappa^2}{l^2 k^4(k^2+\kappa^2)}~
[{\rm G}(ka,kR;\alpha)]^2, \label{vindk-brush}
\end{equation}
whose Fourier transform is
\begin{equation}
v_{\rm ind}(r)~=~-\frac{2Z^2e^2\kappa^2}{\pi\epsilon l^2 r}
~\int_0^{\infty}{\rm d}k\,\frac{\sin(kr)}{k^3(k^2+\kappa^2)} 
~[{\rm G}(ka,kR;\alpha)]^2. \label{vindr-brush}
\end{equation}
For nonoverlapping brushes, Eq.~(\ref{vindr-brush}) can be reduced 
to the analytical form
\begin{equation}
v_{\rm ind}(r)~=~-\frac{Z^2e^2}{\epsilon
r}~+~\frac{Z^2e^2}{\epsilon}~\left[\frac{{\rm S}(\kappa a,\kappa
R;\alpha)}{\kappa l }\right]^2~\frac{e^{-\kappa r}}{r}, \qquad
r>2R. \label{vindr>2R}
\end{equation}
After adding to Eq.~(\ref{vindr>2R}) the bare Coulomb potential 
between the spherical macroions [Eq.~(\ref{vmm})], 
the residual effective pair interaction is 
\begin{equation}
v_{\rm eff}(r)~=~\frac{Z^2e^2}{\epsilon}~\left[\frac{{\rm
S}(\kappa a,\kappa R;\alpha)}{\kappa l
}\right]^2~\frac{e^{-\kappa r}}{r}, \qquad r>2R. \label{veffr>2R}
\end{equation}
Thus, within the coarse-grained PE brush model and at the level of 
linear response theory, nonoverlapping PE brushes are predicted to interact 
via an effective Yukawa pair potential of the same screened-Coulomb form 
as the long-range limit of the DLVO potential~\cite{DLVO} for charged colloids.
This result is consistent with previous linear response results for 
charged hard spheres~\cite{Denton99,Denton00}, 
which interact via the DLVO effective pair potential
\begin{equation}
v_{\rm eff}(r)~=~\frac{Z^2e^2}{\epsilon}~\left(\frac{e^{\kappa R}}
{1+\kappa R}\right)^2~\frac{e^{-\kappa r}}{r}, \qquad r>2R,
\label{vDLVO}
\end{equation}
and for PE stars~\cite{Denton03}, which interact via 
\begin{equation}
v_{\rm eff}(r)~=~\frac{Z^2e^2}{\epsilon}~\left[\frac{{\rm shi}
(\kappa R)}{\kappa R}\right]^2~\frac{e^{-\kappa r}}{r}, \qquad r>2R.
\label{vstar}
\end{equation}
Note that the screening constant, $\kappa$, in the pair potential 
depends on the total density of microions -- inside and outside 
of the brushes -- since all microions respond to the macroion charge. 
We do not consider here overlapping brushes, in which case steric 
interactions between chains also should be included~\cite{Likos02}.

Finally, the volume energy is obtained from Eqs.~(\ref{E0}), 
(\ref{vmckbrush}), (\ref{vindk-brush}), and (\ref{vindr-brush}), as
\begin{eqnarray}
E_0~&=&~F_{\rm OCP}~-~N_m\frac{Z^2e^2\kappa^2}{\pi\epsilon l^{2}}
\int_0^{\infty}{\rm d}k\,\frac{[{\rm G}(ka,kR;\alpha)]^2}
{k^2(k^2+\kappa^2)}~-~(N_+-N_-)\frac{k_BT}{2}.
\label{E0-star}
\end{eqnarray}
Assuming weakly-coupled microion plasmas, the OCP free energy 
is well approximated by its ideal-gas limit:
\begin{equation}
F_{\rm OCP}~=~N_+[\ln(n_+\Lambda_+^3)-1]~+~N_-[\ln(n_-\Lambda_-^3)-1],
\label{FOCP}
\end{equation}
where $\Lambda_{\pm}$ are the thermal de Broglie wavelengths of the
positive/negative microions.  
The physical interpretation of the volume energy is straightforward.
The first term on the right side of Eq.~(\ref{E0-star}) represents 
the entropy of free microions, the second term the electrostatic 
energy of microion-macroion interactions, and the third term 
accounts for the background substraction.  
If the macroion valence $Z$ is allowed to vary with concentration 
({\it e.g.}, through counterion condensation), then $E_0$ should
be supplemented by the macroion self energy.
We emphasize that, because of its dependence on the average macroion 
concentration, the volume energy has the potential to influence 
thermodynamic phase behavior.  As a check of the present results, 
it can be shown that in the two limiting cases of vanishing PE shell 
thickness ($l\to 0$, with Z fixed) and, independently, vanishing 
hard core diameter ($a\to 0$) all analytical results reduce to 
those given in refs.~\cite{Denton99,Denton00} and
\cite{Denton03}, respectively.

\section{Numerical Results and Discussion}\label{Numerical Results}

\begin{figure}
\includegraphics[width=0.6\columnwidth]{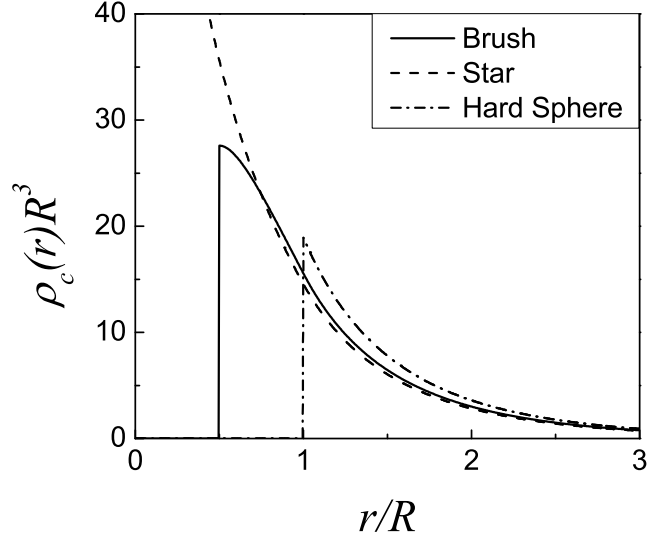}
\caption{\label{rhoc3in1} Counterion number density profiles of 
three types of spherical macroion of outer radius $R=50$ nm,
valence $Z=500$, and reduced number density $n_mR^3=0.01$ in water 
at room temperature ($\lambda_B=0.714~{\rm nm}$):
PE brush-coated macroion [solid curve from Eq.~(\ref{rhocrbrushfull})], 
PE star [dashed curve from Eq.~(20) of ref.~\cite{Denton03}], 
and charged hard sphere [dot-dashed curve from Eq.~(32) of 
ref.~\cite{Denton99}].  For the brush-coated macroion, the hard-core 
radius is $a=25$ nm and the PE shell thickness is $l=25$ nm.}
\end{figure}

\begin{figure}
\includegraphics[width=0.6\columnwidth]{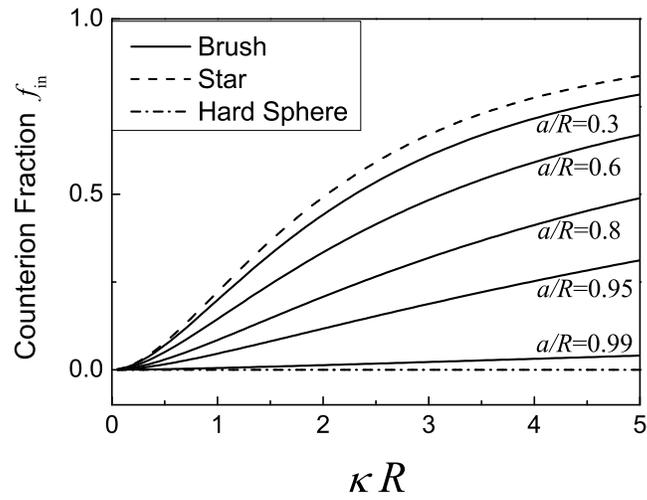}
\caption{\label{ncbrush} Fraction of counterions [from
Eq.~(\ref{ncinside-brush})] trapped inside PE brush as a function of
the dimensionless parameter $\kappa R$ (ratio of outer radius
to Debye screening length) for several values of $a/R$
(ratio of core radius to outer radius).}
\end{figure}

To illustrate applications of the theory developed above, we present 
numerical results for the case of monovalent counterions ($z=1$) 
in aqueous suspensions at room temperature ($\lambda_B=0.714$ nm).  
Figure \ref{rhoc3in1} shows the predicted counterion profiles around 
three different types of macroion, all of the same outer radius $R=50$ nm, 
valence $Z=500$, and reduced number density $n_mR^3=0.01$, for a 
salt-free suspension.  The chosen valence is within the upper limit 
suggested by charge renormalization theory~\cite{Alexander84} for
this size of macroion: $Z<{\cal O}(10)R/\lambda_B$.
For a star macroion, the counterion density diverges logarithmically 
towards the center~\cite{Denton03}, while for brush-coated and
bare hard-sphere macroions the counterion densities remain finite.
Figure \ref{ncbrush} displays the corresponding internal counterion 
fraction, {\it i.e.}, fractional counterion penetration, 
as a function of $\kappa R$.  
For fixed ratio of hard-core radius to outer radius, $a/R$, 
the internal counterion fraction increases monotonically with 
$\kappa R$, reflecting increasing permeability of the macroions to
counterions with decreasing screening length ({\it e.g.}, increasing
salt concentration).  On the other hand, when $\kappa R$ is fixed, 
the counterion penetration decreases upon thinning of the PE brush
(increasing $a/R$).  In the limit of vanishing brush thickness
($l/R\to 0$, $a/R\to 1$), Eq.~(\ref{ncinside-brush}) reduces to
\begin{equation}
f_{\rm in}(l \rightarrow 0)=\frac{\kappa a}{\kappa a + 1} \kappa
l+O(l^2). \label{fin_L_0}
\end{equation}
Counterions are predicted to penetrate PE brush-coated macroions
less efficiently than stars.

\begin{figure}
\includegraphics[width=0.6\columnwidth]{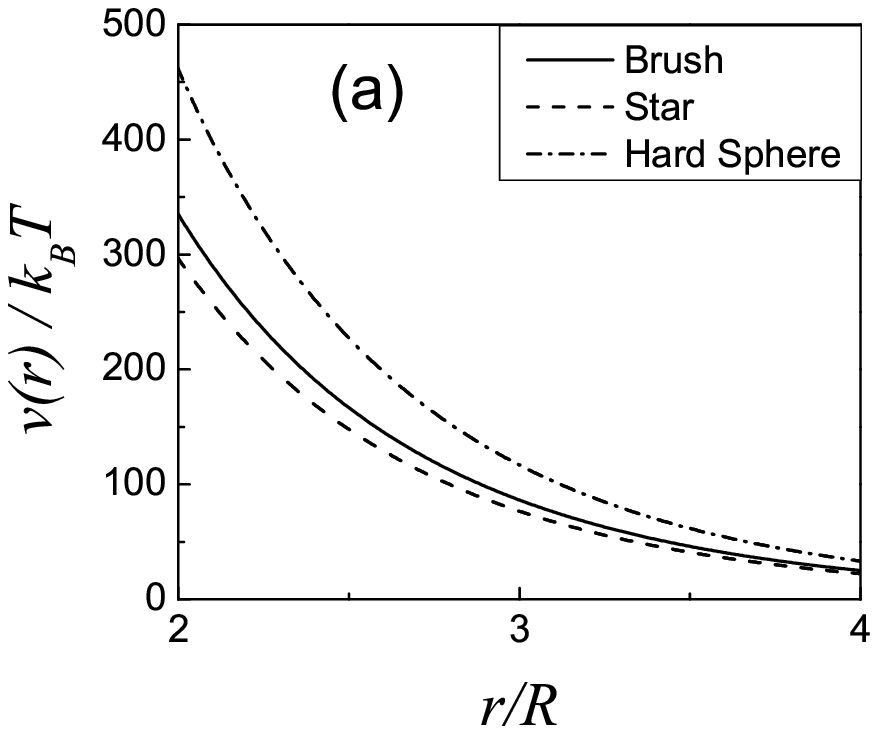}
\includegraphics[width=0.6\columnwidth]{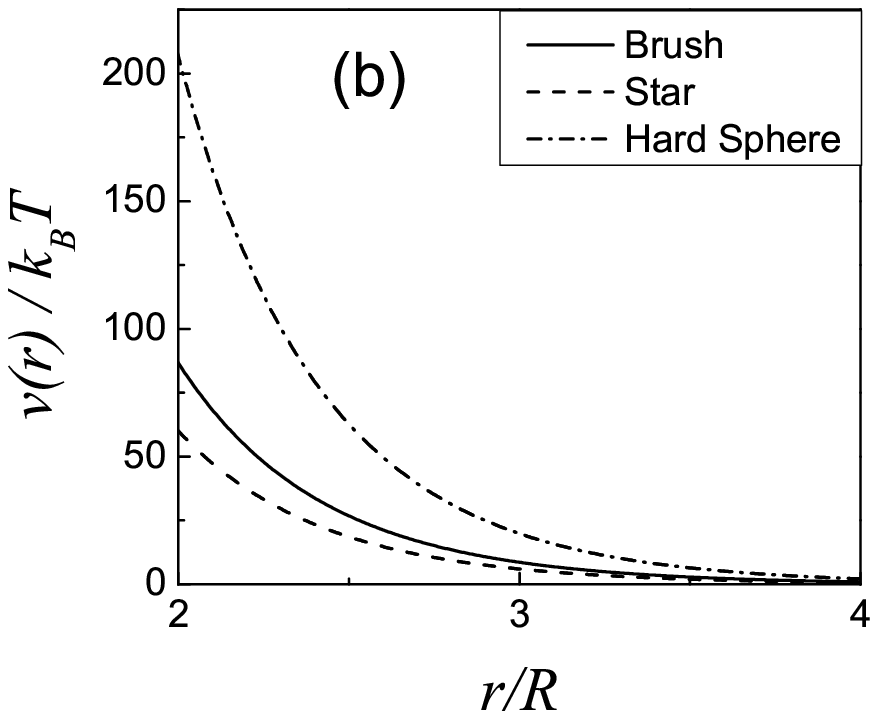}
\caption{\label{veff3in1} Effective electrostatic interactions
between pairs of nonoverlapping macroions of outer radius $R=50$ nm,
valence $Z=500$, and reduced number density $n_mR^3=0.01$ in 
room temperature water ($\lambda_B=0.714~{\rm nm}$) at 
salt concentrations (a) $c_s=0$ mol/l ($\kappa R\simeq 0.95$) and 
(b) $c_s=100$ $\mu$mol/l ($\kappa R\simeq 1.9$):
PE brush-coated spherical macroions [solid curves from 
Eq.~(\ref{veffr>2R})]; PE stars [dashed curves from Eq.~(\ref{vstar})]; 
and charged hard spheres [dot-dashed curves from Eq.~(\ref{vDLVO})].}
\end{figure}

\begin{figure}
\includegraphics[width=0.6\columnwidth]{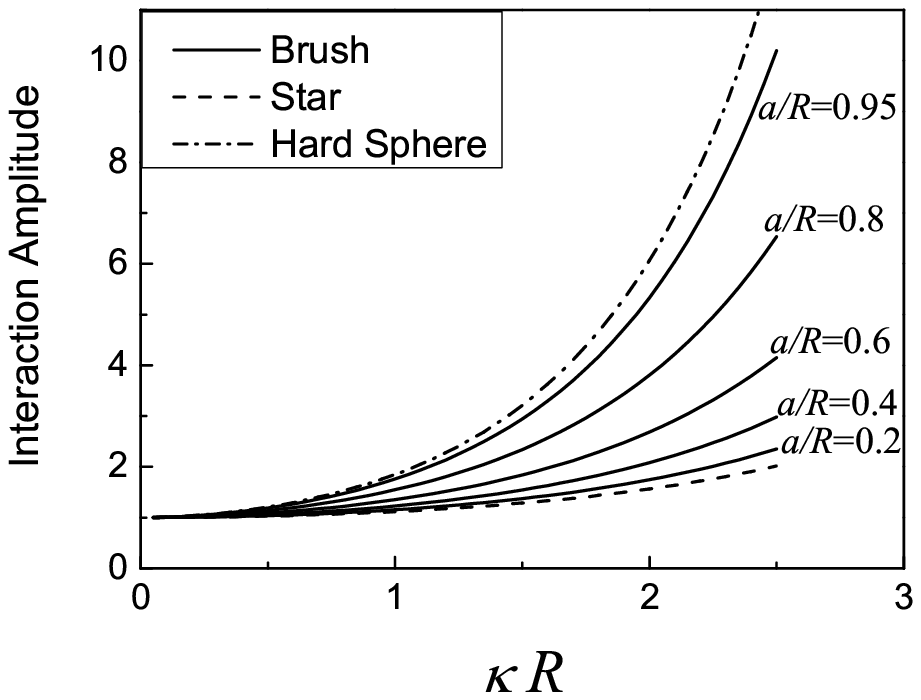}
\caption{\label{amp3in1} Amplitude of Yukawa effective
electrostatic interactions between pairs of nonoverlapping
brushes, stars and charged hard spheres vs. Debye screening constant,
normalized to unity at $\kappa R=0$
[from Eqs.~(\ref{veffr>2R})-(\ref{vstar})].}
\end{figure}

Penetration of macroions by counterions can strongly influence screening 
of bare Coulomb interactions.  Thus, effective pair interactions 
between brush-coated macroions depend sensitively on the thickness
of the PE brush.  To illustrate, 
Fig.~\ref{veff3in1} shows the effective pair potential for the 
same three macroion types as in Figs.~\ref{rhoc3in1} and \ref{ncbrush} 
and for two salt concentrations, $c_s=0$ M and $c_s=100$ $\mu$M,
corresponding to different Debye screening constants $\kappa$.
For identical system parameters, the strength of the Yukawa pair 
interaction for nonoverlapping brush macroions is intermediate 
between that for hard-sphere and star macroions.  Figure~\ref{amp3in1} 
compares the dependence of the macroion-size-dependent amplitude of
$v_{\rm eff}(r)$, $r>2R$, on Debye screening constant for the three 
macroion types.  The amplitude increases with $\kappa R$ for fixed ratio 
of hard-core to outer radius, while for fixed $\kappa R$
the amplitude increases from the star limit to the hard-sphere limit 
as the PE brush thins to infinitesimal thickness ($a/R\to 1$).

\section{Conclusions}\label{Conclusions}

Summarizing, polyelectrolyte-coated colloids provide a valuable 
conceptual bridge between charged colloids and polyelectrolytes.  
In this paper, linear response theory is applied to bulk suspensions 
of spherical colloidal particles coated with PE brushes.  
Assuming stiff, radially stretched PE chains, we model each brush as 
a spherically symmetric shell of continuously distributed charge, 
the charge density varying with radial distance $r$ as $1/r^2$.
By formally integrating out the microion degrees of freedom,
the Hamiltonian of the macroion-microion mixture is mapped onto 
the effective Hamiltonian of an equivalent one-component system.
Predictions of the theory include microion density profiles, 
effective electrostatic interactions between pairs of 
(nonoverlapping) macroions, and a state-dependent one-body 
volume energy, which contributes to the total free energy.
The theory presented here may provide a practical guide for 
choosing system parameters to achieve desired interactions.

The main conclusions of this study are: 
(1) Trapping of counterions inside a spherical PE brush is highly
sensitive to variations in the core radius, brush thickness, and 
Debye screening length of the solution.  
For fixed ratio of core to outer radius, the fraction of trapped 
counterions increases monotonically with increasing outer radius 
or decreasing screening length.
For fixed ratio of outer radius to screening length, the fraction 
of trapped counterions decreases monotonically from a maximum 
in the limit of vanishing core radius (PE star macroion) to zero 
in the limit of vanishing shell thickness (hard-sphere macroion). 
(2) Within the linear response approximation, the effective pair 
interaction between nonoverlapping macroions has a Yukawa 
(screened-Coulomb) form.
(3) By varying core radius and brush thickness, effective interactions 
between PE brush-coated macroions can be tuned -- in both amplitude 
and range -- between interactions for hard-sphere and star macroions.  
For fixed ratio of core radius to outer radius, the amplitude of the 
pair interaction increases monotonically with increasing outer radius 
or decreasing screening length, while
for fixed ratio of outer radius to screening length, the amplitude 
increases monotonically from the star-limit to the hard-sphere limit. 
The range of the pair interaction, governed by the Debye screening 
length, depends on the hard-core volume fraction
and so can be varied by adjusting the core radius.

The range of validity of the coarse-grained model and linearized theory 
studied here, and the accuracy of the predicted Yukawa form of effective 
pair interaction, including amplitude and range, could be directly tested 
by future simulations of more explicit models of PE-grafted colloids.
Our purely electrostatic model can be augmented by chain elasticity and 
entropy -- essential for describing overlapping PE shells~\cite{Likos02}. 
The mean-field linear response theory can be refined to incorporate 
nonlinear microion response~\cite{Denton04} and microion correlations,
beyond the random phase approximation.  The theory also can be easily 
adapted to other macroion types, such as core-shell 
microgels~\cite{Denton03,Hellweg04}.
Future work will explore thermodynamic phase behavior,
which we anticipate to be quite rich and tunable between that of 
charge-stabilized colloidal suspensions and polyelectrolyte solutions.

\begin{acknowledgments}
This work was supported by the National Science Foundation under
Grant Nos.~DMR-0204020 and EPS-0132289.
\end{acknowledgments}

\clearpage

\begin{references}
\bibitem{note1}
The background energy, which is infinite for point counterions, 
formally cancels corresponding infinities in the counterion and
macroion-counterion Hamiltonians [Eqs.~(\ref{Hc}) and (\ref{Hmc1})].
\end{references}
\end{document}